\documentclass[a4paper,twoside,11pt]{Latex/Classes/PhDthesisPSnPDF}
\usepackage[T1]{fontenc}
\usepackage[utf8]{inputenc}

\newcommand{\figuremacroW}[4]{
	\begin{figure}[htbp]
		\centering
		\includegraphics[width=#4\textwidth]{#1}
		\caption[#2]{\textbf{#2} - #3}
		\label{#1}
	\end{figure}
}

\usepackage{graphicx}
\usepackage{etoolbox}
\def\@degree{\@latex@error{No \noexpand\degree given}\@ehc}

\begin{document}



\begin{titlepage}
	\centering
	{\scshape\LARGE INSTITUTO DE ASTROFISICA DE CANARIAS - UNIVERSIDAD DE LA LAGUNA \par}
	\vspace{1cm}
	{\scshape\Large MASTER OF ASTRONOMY\par}
	\vspace{1.5cm}
	{\huge\bfseries Ageing of a space-based CCD: photometric performance development of the low Earth orbiting detectors of the CoRoT mission\par}
	\vspace{2cm}
	{\Large \bfseries Ruben Asensio--Torres\par}
	\vfill
	supervised by\par
	Dr.~Hans J. \textsc{Deeg}

	\vfill

	{\large January 2015}
\end{titlepage}

       


\frontmatter

\cleardoublepage


\begin{abstracts}        

In this thesis we have analysed the time evolution of the photometric precision achieved by the space-based exoplanet-hunting mission \textit{CoRoT} during its flight phase (2007-2012). This study of the noise level of CoRoT light curves has been based on a previous paper by Aigrain et al. 2009, where they found \textit{a gradual degradation of the photometric performance over time} for the first 14 months of data. \\

A pre-processing of the light curves is necessary to remove unrelated flux variations from the target, such as cosmic rays hitting the CCDs or when the satellite crosses the Earth's shadow. These effects produce light curves with a very high number of outliers that invalidate the measurements performed by CoRoT. We obtain the scatter in magnitudes from the corrected light curves for point-to-point observations and on typical transit time scales for CoRoT targets, in this case 2h. Here we have analysed the anti-center runs IRa01 (2007), LRa01 (2008), LRa03 (2010) and LRa06 (2012). The two first runs were studied by Aigrain as well, so we are able to compare our results. The two last runs allowed us to evaluate the trend of photometric degradation over more than 5 years\\

We obtain low observational point-to-point noise, although a factor 3 bigger than the source photon noise. As expected, this noise increases versus magnitude as does the photon noise for fainter stars. We find here a difference with the results reported by Aigrain, who found a steeper slope for the point-to-point noise versus brightness. We find effects showing the ageing of the CoRoT CCDs.  On 2h time scales we notice a receding photometric performance, with a noise increase of about 2.1 times across the four analysed runs, corresponding to a $\sim$15$\%$ increase per year. Correlated noise becomes more important than white uncorrelated noise for the two last studied runs, LRa03 and LRa06. The strongest degradation, however, occurs during the first year of operations, with a 30$\%$ noise increase, opening up a two-ageing-timescales scenario. This higher level of scatter is probably a consequence of the increment of \textit{hot pixels} hitting the detector.  Finally, by means of the measurement of the photometric zero point we verify that there is no significant change in the instrument efficiency over the flight phase.\\

\end{abstracts}



\setcounter{secnumdepth}{3} 
\setcounter{tocdepth}{3}    
\tableofcontents            


\mainmatter

\renewcommand{\chaptername}{} 


\chapter{INTRODUCTION}

\ifpdf
    \graphicspath{{1_introduction/figures/PNG/}{1_introduction/figures/PDF/}{1_introduction/figures/}}
\else
    \graphicspath{{1_introduction/figures/EPS/}{1_introduction/figures/}}
\fi


The CoRoT (Convection, Rotation and planetary Transits) satellite was the first instrument dedicated to finding extrasolar planets from space \cite{deleuil01}. It was launched into a polar inertial circular orbit in December 2006 and started the observations on February 2nd, 2007. Ultra-high precision photometry was used to achieve the two scientific goals of the mission: the study of stellar interiors by performing asteroseismology and the detection and characterization of extrasolar planets by the transit method, measuring the decrease of the received flux when the star, the planet and the satellite are aligned.\\

\section{Charge-Coupled Devices}

Astronomers began to adopt CCDs (Charge-Coupled Devices) in the 1970s as detectors of radiation. Their ability to integrate over long time intervals, their electronic readout,  high quantum efficiency and large dynamic range ended with the photographic plate era.\\

The mechanism of a CCD is based on the photoelectric effect. Electrons produced by incident photons on a semiconductor, usually silicon, are trapped as packets of charge in potential wells produced by a localized electric field. The fundamental element of a CCD (i.e. a pixel) is shown in Figure \ref{ccd.png}. It is composed of a semiconductor substrate and over it there is an oxide or other dielectric layer. The function of this layer is to insulate an electrode from the semiconductor. By doping the silicon with an acceptor, typically boron, holes become  the predominant carriers in the semiconductor, forming p-type silicon. When the electrode is held at a positive voltage, the holes are repulsed from it creating a hole depletion region, while the charged-negative electrons are attracted to the storage region close to the electrode. The incident photons produce electron-hole pairs in this depletion region that are split apart, accumulating the electrons in the storage region. However, as a result of thermal fluctuations, electrons in the storage regions may appear spontaneously, creating what is called dark current. This thermal process can be reduced by cooling the detector at as lower temperature as is consistent with good charge transfer efficiency (Leach 1987\cite{leach01}; Leach 1980)\cite{leach02}.\\

\figuremacroW{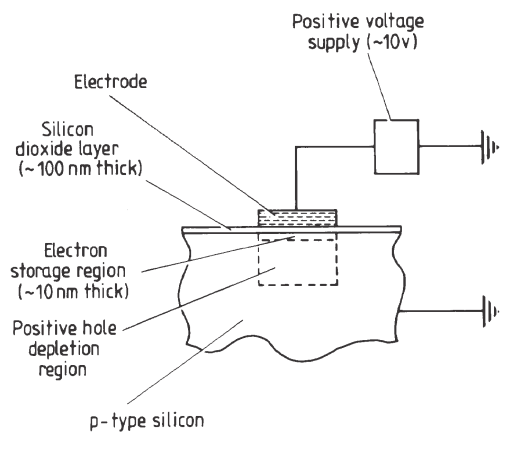}{\footnotesize{Basic unit of a CCD.}}{\footnotesize{Taken from \cite{kitchin01}}}{0.5}

	An individual pixel unit is not very useful by itself. CCDs can only be of utility as an array of pixels. Each one of them will detect a charge that is proportional to its illuminating intensity. In order for the charge to be read out, the electrons stored in each pixel are moved physically to the adjacent electrode by sequentially changing the voltage of the electrodes (Kitchin 2003). This charge transfer is not perfect and some electrons might be lost, although with a good design the efficiency of the charge transfer may be almost perfect. Ultimately, the charge will be brought to an output electrode where the charge is determined using readout electronics. The digitisation of the signal starts by a pre-amplification stage and is followed by a signal processing (see Figure \ref{readout.png}). An A/D converter converts the electrons to the detected ADUs (Analogue-to-Digital Units). The gain factor gives the conversion of digital counts to electrons. This process of reading the amount of charge in a pixel adds the so-called read-noise. Finally, CCDs have different electric sensitivity to the incident radiation depending on its wavelength. The quantum efficiency of the detector measures the proportion of photons hitting the detector that is converted into electrons. \\

\figuremacroW{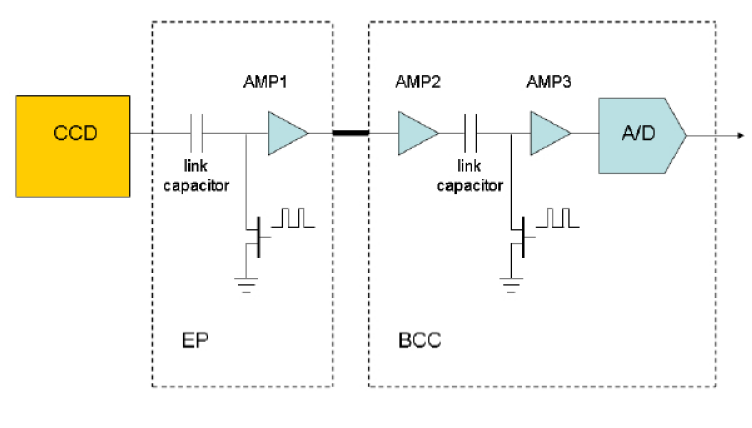}{\footnotesize{CoRoT readout electronics}}{\footnotesize{A pre-amplification stage (EP) is followd by a signal processing (BCC). Taken from\cite{auvergne01}}}{0.6}

	The described sources of noise affect the observed signal. The total noise will be given by:\\
\begin{equation}
Total Noise = \sqrt{ RN^{2} + DCN^{2} + BPN^{2} + SPN^{2} }
\end{equation}\\
where $RN$ is the read noise, $DCN$ is the dark current noise, $BPN$ is the background photon noise and $SPN$ is the source photon noise. The photon noise is associated with the particle nature of light and has a Poisson distribution. These noise sources are independent so they get added in quadrature.\\

The signal-to-noise ratio, on the other hand, measures the quality of a CCD measurement (or exposure):\\

\begin{equation}
\frac{S}{N} = \frac{tP_{s}}{\sqrt{qN_{r}^{2}+ qi_{d}t+qatP_{back}+tP_{s}}}
\end{equation}
where $q$ is the number of pixels in the aperture, $t$ is the integration time, $a$ is the are on sky of one pixel, P$_{s}$ is the photons per second from the source, P$_{back}$ the background photons per second per square arcsecond and $Nr$ the read noise.

\section{CoRoT: instrument and performance}

CoRoT's afocal telescope is accommodated in a 2 stage baffle, blocking sunlight reflected by the Earth. Two mirrors bring the observed light to a focal box which contains 4 CCDs, one pair dedicated to the exoplanetary channel and one to the asteroseismology channel. This architecture achieves a field of view of 2.7 $\times$ 3.05 degrees and provides a collecting surface able to obtain a high enough signal to noise ratio for exoplanet observation on stars with R magnitude between 11.5 and 16 \cite{auvergne01}.\\

The four E2V 2048 $\times$ 2048 pixel, back-illuminated CCDs with anti-reflection coating are located in the focal plane and shielded against cosmic rays by a 10mm thick aluminium shelter. Each pixel is 13.5 $\times$ 13.5 $\mu$m$^{2}$ in size, corresponding to an angular size of 2.32'' $\times$ 2.32''  in the sky.  The flatness of the image section is about 20$\mu$m peak-to-peak \cite{bernardi01}. To ensure stability of the photometric signal of CoRoT, the detectors are cooled at -36ºC  and a telescope stability better than 0.3K on the orbital time scale of 46 minutes and better than 1K over a long run must be maintained. This gives a dark current rate of 0.6 electrons/pixel/sec. The digitisation of the signal on CoRoT is performed for each half CCD and the factor of conversion of ADUs to electrons , the so-called gain, is 2 electrons/ADUs \cite{auvergne01}. The quantum efficiency limits the detection bandwidth in the wavelength interval 250 – 1000 nm, which peaks at around 650 nm. A prism in front of the two CCDs for exoplanetary search disperses the light and forms a small spectrum of each target. The flux is computed using a mask selected among 255 templates, which encompasses the PSF. CoRoT PSFs, with typical sizes of 35'' $\times$ 23'' in the exoplanet channel, are slightly larger because the CCDs are defocused to avoid saturation of the brightest stars \cite{deleuil02}. However, this fact does not harm the photometric performance, except for the addition of further read-noise and photon-noise from potential background stars.\\
	
	CoRoT's viewing zone is in the equatorial direction in order to avoid occultations by the Earth, nor to point close to the Sun. For the same reason, the satellite is reversed and pointed in the opposite direction every 6 months. In this way, CoRoT is constrained to observe within two fields in the galactic plane, known as the $CoRoT eyes$. These zones are 10$^{\circ}$  of radius \cite{auvergne01} and are centered at right ascension 6h50min, towards the galactic anit-centre, and 18h50m, towards the galactic centre, both being at 0$^{\circ}$ declination.\\

	The observing runs consist of one Long Run of 150 days followed by one or more Short runs of 20 or 40 days. The initial run was the exception, lasting only 75 days. The runs are referred as IR, SR or LR (for initial, short and long run) followed by $a$ or $c$ (towards the galactic anti-centre or centre, respectively) and two digits for its numbering (i.e. IRa01, LRa06, SRc01... ).  Table \ref{field_description} shows the runs performed by CoRoT. A failure of Data Processing Unit 1 in March 2009 meant that the payload had to operate with only half the field of view for both the exoplanet search and the asteroseismology channel. CoRoT stopped providing data on 2012-11-02 when the second photometric chain failed to communicate with the spacecraft computer. \\

\begin{table}
\centering
\begin{tabular}{cccccccc} 

\textbf{\footnotesize{Run ID}} & \bf \footnotesize{Run Start} & \bf \footnotesize{Run End} & \bf $\alpha$ ($^{\circ}$) & \bf $\delta$ ($^{\circ}$) & \bf $\phi$ ($^{\circ}$) & \textbf{\footnotesize{D(d)}}& \textbf{\footnotesize{TS(yr)}}  \\ 
\hline 

\textbf{IRa01} & \textbf{2007-01-18} & \textbf{2007-04-03} & \textbf{102.604} & \textbf{-1.7} & \textbf{9.596}&\textbf{75}&\textbf{0.16}\\
SRc01 &2007-04-03&2007-05-09&284.593&3.08&5.483&36&0.31\\
LRc01 &2007-05-09&2007-10-15&290.889&0.434&24.244&159&0.8\\
\textbf{LRa01}&\textbf{2007-10-15}&\textbf{2008-03-03}&\textbf{101.664}&\textbf{-0.2}&\textbf{1.916}&\textbf{140}&\textbf{0.99}\\
SRa01&2008-03-03&2008-03-31&101.037&9.022&2.316&28&1.22\\
LRc02&2008-03-31&2008-09-08&279.664&6.4&16.723&161&1.48\\
SRc02&2008-09-08&2008-10-06&284.108&-2.864&-14.636&28&1.74\\
SRa02&2008-10-06&2008-11-12&97.555&5.667&-25.363&37&1.82\\
LRa02&2008-11-12&2009-03-30&103.519&-4.38&5.996&138&2.06\\
LRc03&2009-03-30&2009-07-02&277.470&-7.245&16.243&94&2.38\\
LRc04&2009-07-02&2009-09-30&277.72&	8.02&6.563&90&2.63\\
\textbf{LRa03}&\textbf{2009-09-30}&\textbf{2010-03-01}&\textbf{93.745}&	\textbf{5.503}&\textbf{-3.843}&\textbf{152}&\textbf{2.97}\\
SRa03&2010-03-01&2010-04-02&98.401&0.99&2.156&32&3.22\\
LRc05&2010-04-02&2010-07-05&279&3.66&6.767&94&3.39\\
LRc06&2010-07-05&2010-09-27&278.951&4.35&16.847&84&3.63\\
LRa04&2010-09-27&2010-12-16&92.571&6.940&5.872&80&3.86\\
LRa05&2010-12-16&2011-04-05&91.930&7.944&-23.487&110&4.12\\
LRc07&2011-04-05&2011-06-30&277.604&6.29&23.967&86&4.39\\
LRc08&2011-07-06&2011-09-30&277.144&5.6&-16.032&86&4.64\\
SRa04&2011-09-30&2011-11-28&96.172&-3.842&1.792&59&4.84\\
SRa05&2011-11-29&2012-01-09&101.124&10.07&-18.607&41&4.98\\
\textbf{LRa06}&\textbf{2012-01-10}&\textbf{2012-03-29}&\textbf{101.545}&\textbf{0.22}&\textbf{16.032}&\textbf{79}&\textbf{5.14}\\
LRc09&2012-04-10&2012-07-05&288.224&-3.18&-1.152&86&5.40\\
LRc10&2012-07-06&2012-10-01&277.743&7.28&-9.632&87&5.64\\
LRa07&2012-10-01&2012-11-02&96.867&5.188&15.232&98&5.89\\
\end{tabular}
\caption[table]{\textbf{Field Descriptions} - Run ID, center-positions ($\alpha$ = right ascension and $\delta$ declination), roll angles ($\phi$) and duration (D) in days of the CoRoT runs. Time in space (TS) shows the time since launch on 28 Dec 2006, taken from the middle of the run. In bold, runs studied in this work.}
\label{field_description} 
\end{table}
	The science data is perturbed by the closeness of the satellite to the Earth. These effects must be taken into account and corrected to carry out an analysis of CoRoT's performance. To start with, the relative position between the Sun, the Earth and the satellite makes CoRoT to go in and out the Earth's shadow, causing voltage and temperature fluctuations. These eclipses are longer at the beginning and at the end of a run.  On the other hand, the Van Allen belts are regions of the Earth’s magnetosphere where energetic charged particles are magnetically trapped. The outer belt is populated by electrons with energies of keV to MeV, while the inner belt consists of an intense radiation of energetic protons up to a few GeV \cite{adriani01}. CoRoT spends 7\% of the time in the region where the inner belt makes its closest approach to the Earth's surface \cite{pinheiro01}, known as SAA (South Atlantic Anomaly), located over South America. As a consequence, the CCD is hit by more energetic particles than usual when crossing the SAA.\\

	A pre-launch photometric precision of 7 $\times$10$^{-4}$ for a V=15.5 magnitude and one hour of integration was specified in order to be able to detect telluric planets \cite{auvergne01}. Here we analyze the evolution of the photometric performance for CoRoT, studying the noise of its light curves during the flight phase. CCD's performance can be altered by a variety of effects such as hot pixel events, changes in its spectral response or saturation level. Our study will be based on Aigrain et al. (2009)\cite{aigrain01}, from now on A09, where evidence of slight degradation over time was found for the first 14 months of the satellite's operations.\\

\chapter{DATA ANALYSIS METHODOLOGY}

\ifpdf
    \graphicspath{{2/figures/PNG/}{2/figures/PDF/}{2/figures/}}
\else
    \graphicspath{{2figures/EPS/}{2/figures/}}
\fi


In this section we describe how we ingest, prepare and treat the data for the analysis.

\section{Data Collection}

The CoRoT science-grade N2 data is released in FITS (Flexible Image Transport System) files and contains information on the light curves in the form of binary tables and some auxiliary information put in the header of the light curves files.  Although we are also provided with the astereoseismology data, we will only make use of the exoplanet data channel. The Python programming language has been used for the data collection, the complete processing of the data and the final representation of the results.\\

The date, flux and \textit{status} (flags indicating the status of the measurement) of each exposure are read out from the binary tables. While the integration time is 32 sec, usually 16 observations are added on board covering 512 sec before being downloaded. However, this is not always the case. For specific stars where it is more likely to find transits, like bright dwarfs, or for stars for which a potential transit "alarm" was found, the 32 sec data is preserved.\\
 
For stars brighter than magnitude 15, the flux is given in 3-colour photometry; red, green and blue. These are the chromatic light curves. The colour designation in CoRoT is rather different from standard astronomical filters.  The two CCDs dedicated to exoplanetary search are behind a prism that disperses the light and produces a spectrum for each star. Three subdivisions of pixels are separated within a photometric mask, whose flux is recorded independently for the three colours designated blue, green and red. As it is not possible to split a pixel in half, and due to the use of over 200 different aperture masks (adapted to the star's spectral energy distribution, position on CCD and neighboring stars location), each star has a slightly different distribution of its energy in each colour. Monochromatic light curves provide the total \textit{white} flux. We will focus on the white flux, so when a chromatic flux is provided, we add up the fluxes of the three colors.\\

From the header of the light curves we get the red, green and blue Johnson -Kron-Cousins standard magnitudes (as used by Landolt 1992) of each target, taken from the Exodat catalogue \cite{deleuil01}, which were obtained from 4-colour photometry from the INT telescope on La Palma. We use the Python function \textit{getheader} inside the module \textit{pyfits} to get them.\\

Our study will be based on runs IRa01, LRa01, LRa03 and LRa06. This selection was based on several reasons: first, it allows us to match our results with those obtained by Aigrain, as runs IRa01 and LRa01 are present in both studies; further, these runs are split in time in a more or less regular basis, which makes it possible to study the photometric performance of CoRoT's CCD over time (see right-most column in Table \ref{field_description}). Finally, the obtained magnitudes from CoRoT when observing at the galactic center are less accurate than those from the anticenter because of light contamination from other sources other than the target star, so for consistency, we selected only anti-center runs. Lastly, these four runs all had complete coverage by the INT-derived photometry, which was not the case for all runs (USNO catalogue photometry was used in some others, which has much less precision)\\

\section{Noise Reduction}

Once we have the date, flux and status of every measurement in a light curve, there is noise from known sources, other than the targets, which has to be removed before looking for transits. After this process, we will be able to study the noise present in each light curve over 512 seconds and on transit time scales.\\   

As stated in section 2.1, those measurements affected by these known noise sources have previously been detected and flagged. These flags indicating the status of each measurement are contained in the \textit{status} binary table extension. However, taking into account that the flagging process is not perfect, we will perform a removal of the noise using our own system based on the one described by Aigrain.\\

The most striking feature when looking at the raw light curves is the high number of mostly upwards outliers (see also Figure \ref{51.pdf}). The downwards outliers do not affect the data in a great manner, as they are caused by a short drop in observed flux due to the loss in pointing accuracy when CoRoT experiences eclipses. These flux perturbations are clipped out by means of the flagging extension, but those binned on board to a 512 sec sampling time were not flagged properly by the pipeline (A09). We will let our filtering process to remove most of the affected downward outliers.\\

Some of the data was not rebinned on board to a 512 sec sampling time. Where we find 32 sec exposures, we ourselves rebinned the data in order to have an even 512 sec sampling by taking the median of each 16 exposures. The use of the median minimizes the effect of outliers. In fact, within the same light curve we find often 32 and 512 sec sampling together. Uneven sampling or missing data points might also be present, so if within 512 sec there are less than eight 32 sec exposures, we eliminate the whole 512 sec block.\\

The more prominent upward outliers are caused by energetic particles hitting the CCD when the satellite crosses the  SAA (South Atlantic Anomaly). This higher background produces a high number of upward outliers. The removal of outliers is based on a filtering process, where a mean 5-point boxcar filter is first applied to the data followed by a median filter with a 1-hour baseline. Then, a 3-sigma clipping is performed to reject the detected outliers. This process is iterated until it converges, normally after only a few iterations.\\

The mean 5-point ($\sim$43 minutes if no data points are missing) boxcar filter is applied to the data first. For every data point, the mean value of the four surrounding points (the two points before and after it) and the point itself is calculated and used as a value for a new, smooth list of data points. This new list has fewer outliers, benefiting from the mean standard filter. After this smoothing process, the median filter with 1-hour baseline is applied to the obtained new list. It works as the boxcar filter does, but the median is less affected by the presence of outliers. Again, a new list of data points is obtained. The data is then split into 10 blocks and, for each one, the MAD (Median Absolute Deviation)-estimated standard deviation is calculated from the list. The difference between this list and the original light curve is used for the 3-sigma clipping of the original light curve. The filtering is again repeated until the ratio of the standard deviation of the new list to the standard deviation of the original light curve is higher than 75\%. In consequence, some data points might be missing after each iteration. In order to avoid big gaps, filters will move on to the next point only if more than half of the points that the baseline would contain without clipping were present. If that is not the case, the filter stops and starts over after the gap.\\

Most of the light curves present a downward trend or decay throughout a run, which for Aigrain is a consequence of a pointing drift in the satellite's field of view. We can imagine that, as the pointing drifts away from the observed star,  less flux is observed in the star aperture because we would be getting background photons. However, this motion over the aperture would imply that a star would be detected redder or bluer with time, which is not the case. Deleuil attributed  the flux decrease to a gain decrease. This explanation would require a similar behavior of the readout electronics (amplifier + A/D converter) in both CCDs over time, which is also not very likely. A third solution for this problem could be a decrease in the quantum efficiency of the detector, something that we evaluate in Section 2.4 and 3.3.\\

In order to remove this trend, the \textit{detrend} function in Python is used. The result of a linear least-squares fit to the light curve is subtracted from the clipped light curve, point by point. This is, for every observation we subtract the value of the fit at the time of each data point from the original flux. We come back to original fluxes by adding the median value of the clipped light curve to the detrended light curve.\\

Periodical variations due to rotational modulation and the evolution of active regions on the surface of the stars are expected to be present in the light curves. We use a non-linear filter such as the one used for the 3-sigma clipping to obtain these long-term variations. This time the median filter has a 1-day baseline, though, acting as a high pass filter. The median for the 1-day surrounding points is calculated and subtracted from the detrended light curve. In this way, we remove low frequency periodical variations (longer than 1 day), much longer than the typical CoRoT transit time scale which is about 2h (A09). Again, the median of the light curve was added to recover the original flux values.\\
	
\section{Scatter Estimates}

The detectability of transits will depend on the level of noise of the processed light curve. The corrections described in section 2.2 should have removed any astrophysical signal not related to transits, but the remaining noise level is evaluated by means of the MAD statistic. The MAD is calculated by taking the median of the absolute deviations from the observation's median:

\begin{equation}
MAD = median(|x_{i} - median(X)|)
\end{equation}
where $x_{i}$ represents a value of the dataset and X the set of all the x$_{i}$. As a still important number of outliers might remain in the data after the corrections are applied, the MAD statistic, which is little sensitive to the presence of outliers, becomes a robust measure of the standard deviation $\sigma$. We can take $\sigma$=1.4826$\times$MAD (A09). From the standard deviation, the scatter in magnitudes is derived as \cite{deeg01}:

\begin{equation}
Scatter = -2.5log(1- \frac{\sigma}{median(flux)})
\end{equation}

where $median(flux)$ is the median of the flux values of the processed light curve. In this way, we convert flux in ADUs to magnitudes.\\

	We might assume at this point that the noise present in the light curves is simply uncorrelated white noise. However, some systematics related to a combination of factors, as orbital motion, telescope tracking, satellite jitter and stellar activity appear in the data \cite{pont01}; \cite{mazeh01}). From these sources, the noises present in the observations will be correlated. For a typical CoRoT transit with 2h duration, two components of the noise will be affecting the observations: white random noise and a low-frequency (red) correlated noise (see Figure \ref{rednoise.png}). As a consequence, we can't infer the noise on 2h time scales by scaling the point-to-point noise $\sigma_{pp}$ by the inverse of the square root of the number of data points on a 2h interval to $\frac{\sigma_{pp}}{\sqrt{14}}$, which would be the case if only uncorrelated noise were present.\\

\figuremacroW{rednoise.png}{\footnotesize{Effect of red correlated noise on light curves.}}{\footnotesize{Top figure: light curve with only white noise. Middle figure: light curve with} only red noise. Bottom figure: light curve with both white and red noise. Taken from \cite{pont01}}{0.4}

	The scatter given by equation 2.2 will be evaluated over 512 sec (point-to-point), over 2h (typical transit time scales) and on long time scales (using the light curve obtained after the removal of outliers and before the detrending).

\section{Throughput evolution}

The number of electrons produced in the CCD from a given source depends not only on the flux that the primary mirror receives from the target, but also on the total optical transmission of the instrument (telescope optics, instrument optics, prism...) and the quantum efficiency of the detector. The total transmission plus the detector quantum efficiency determine the fraction of photons reaching the telescope that will be detected. Moreover, as after the amplifier and the analogue-digital conversion what we measure are ADUs, the amplifier gain will also alter the number of counts obtained.\\

	This total throughput can be estimated by comparing CoRoT's measured flux values of real stars with known standard values, in this case given by Exodat. Comparing the results obtained for different runs, we can study the evolution of the telescope's global transmission, quantum efficiency and gain.


\chapter{RESULTS}


\ifpdf
    \graphicspath{{3/figures/PNG/}{3/figures/PDF/}{3/figures/}}
\else
    \graphicspath{{3/figures/EPS/}{3/figures/}}
\fi


\section{Light Curve Correction}

Raw light curves from CoRoT need to be corrected before searching for stellar companions or analysing the level of noise present on transits timescales. Figure \ref{51.pdf} shows the correction methods described in Section 2.2 applied to two light curves from LRa06, taken during the period 2012/01/10 - 2012/03/29 (see Table \ref{field_description}). The yellow light curve is the raw data directly taken from the N2 data binary table. The most striking feature is the high number of upward outliers caused by SAA protons hitting the CCD. About 8 pixels are altered per particle impact, 90\% of which are energetic enough to invalidate the aperture photometry measurement \cite{pinheiro01}. To a lesser extent, downwards outliers are also present as a consequence of the loss of pointing accuracy when traveling in and out of the shadow of the Earth. \\

\figuremacroW{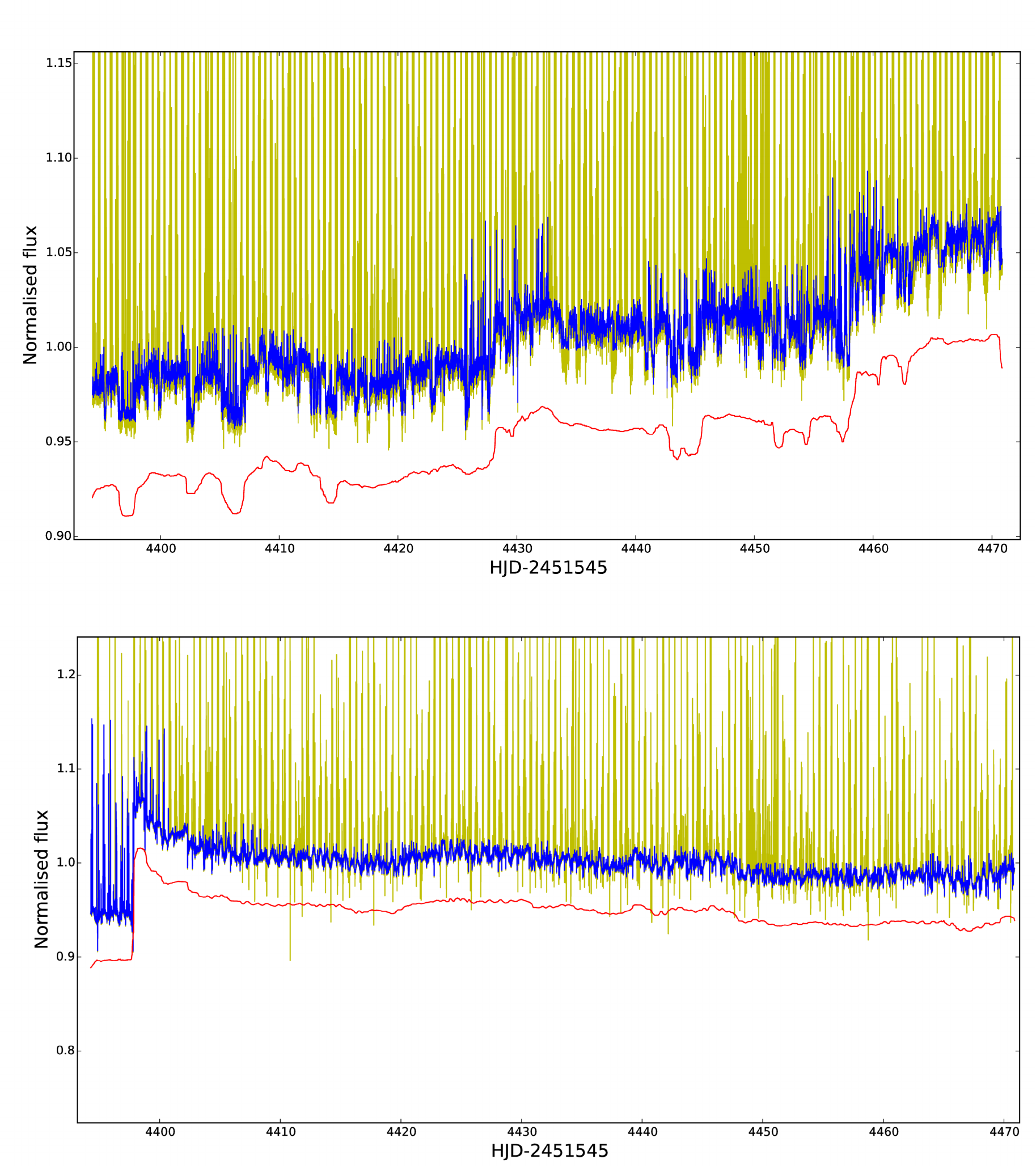}{\small{Correction methods applied to two lightcurves with peculiar features from LRa06}}{\small{Raw light curve shown in yellow. The result of the 3-sigma filtering process is shown in blue and variatons longer than a day are shown in red with a small offset for a good comparison. Top figure: a light curve with a clear upward trend. Bottom figure: a light curve affected by a \textit{hot pixel} event near 4400.}}{1}

In order to remove these outliers, the 3-sigma clipping filtering process described in Section 2.2 is performed until convergence. Most of the light curves present either a decay or an upward trend
of the photometric flux within the aperture (an example of upward linear trend can be seen in Figure \ref{51.pdf}). We perform a detrending process for which the result of a least-squares fit to the light curve is substracted point by point. Finally, variations longer than a day are obtained and removed by applying the filter used for the 3-sigma clipping, this time with a 1-day baseline, and substracting it from the detrended light curve. These long-term variations are clearly seen in red in Figure \ref{51.pdf}. For clarity purposes we show these variations before detrending and with an offset.\\

\figuremacroW{1.pdf}{\small{Result of the corrections applied for a given light curve in LRa06}}{\small{Top figure: raw light curve obtained from the FITS files. Bottom figure: same light curve ready for transit search after all the corrections that we described}}{1}

Some of the light curves contain an effect similar to what we can see in the bottom light curve in Figure \ref{51.pdf}: a jump followed by an exponential decrease that slows down over days and finally stabilizes to a new flux value. This event is due to the atomic displacement in the silicon lattice of the CCD caused by the impact of charged particles. It causes permanent damage that leads to mid-band energy levels, which in turn allows thermal fluctuations to enhance dark current rates. This effect can be very large and localized, with pixels presenting 0h high dark current being called \textit{hot pixels}. The evolution of the intensity of hot pixels is not stable with time, but its time evolution varies due to the arrangement of the silicon defects in more stable configurations. It can disappear after days or even years \cite{pinheiro01}.\\

\section{Scatter on point-to-point and transit timescales}

The level of scatter present in CoRoT light curves sets a threshold on the detectability of potential extrasolar planets orbiting around other stars. After all the corrections and the removal of outliers that we have applied to the raw light curves, it is assumed that only two components remain in the noise: a white-noise component without any temporal corrections and a component with systematic temporal correlations. The light curves' scatter in magnitudes has been evaluated over 512 sec (point-to-point), over transit time scales (2h) and over long time scales (longer than about 1 day).\\

The MAD-evaluated standard deviation on 2h timescales is then composed of these two components as:

\begin{equation}
\sigma_{2h} = \sqrt{\sigma_{uncorrelated}^{2} + \sigma_{correlated}^{2}}
\end{equation}

We have performed the estimation of $\sigma_{2h}$ from the MAD of the median points of each two-hour block. The conversion to scatter in magnitudes on 2h time scales will be obtained empirically by applying Eq (2.2) to $\sigma_{2h}$. The white noise over 2h is obtained as the ratio of the point-to-point noise over the square root of the number of points in a 2h interval, which is 14 for 512 sec exposures (e.g $\sigma_{uncorrelated} = \frac{\sigma_{pp}}{\sqrt{14}}$). The correlated component of the noise is then derived from solving Eq (3.1) for $\sigma_{correlated}$. In the same way, from the light curve before detrending but after the 3-sigma clipping (red one in Figure \ref{51.pdf}), we can evaluate the long term component of the noise by means of the binning of 1-day blocks. \\

\figuremacroW{3.pdf}{\small{Noise level against R-magnitude for IRa01, LRa01, LRa03 and LRa06}}{\small{Black dots represent the point-to-point scatter for each light curve. The solid line is a lower envelope of these points with an approximated global slope of 0.20 (see Table \ref{fits}) obtained from a least-squares fit to the 20\% light curves with less noise. By scaling the solid line down by a factor $\sqrt{14}$ (being 14 the number of 512 sec observations present in a 2h block) we obtain the white uncorrelated noise, which is the lowest dotted line. The theoretical photon source noise is shown as the dashed curve with slope 0.20. Correlated noise on 2h time scales is calculated from Eq (3.1) and represented as green dots. Scatter on longer time scales than a day is shown as red dots. This figure is similar to Figure 2 of A09, except for the colours used.}}{1}

\figuremacroW{4.pdf}{\small{Total two hour scatter against R-magnitude for IRa01, LRa01, LRa03 and LRa06}}{\small{Scatter on 2h time scales for likely dwarf stars is shown as blue dots, while giants are shown in orange. The distinction between likely giants or dwarfs is based on Figure \ref{2.pdf}. The thick solid line, with a slope close to 0.20 (see Table \ref{fits}), is obtained from a least-squares fit to the 20$\%$ dwarf stars with less noise. In the analysed runs, no relevant difference in noise from giant and dwarfs stars is seen, except that the giant star's noise bottoms out at a scatter of 2-3 $\times$ 10$^{-4}$. White uncorrelated noise is shown for comparison as the same dotted lines as in Figure \ref{3.pdf}. This figure is similar to Figure 3 of A09, except for the colours used.}}{1}

\figuremacroW{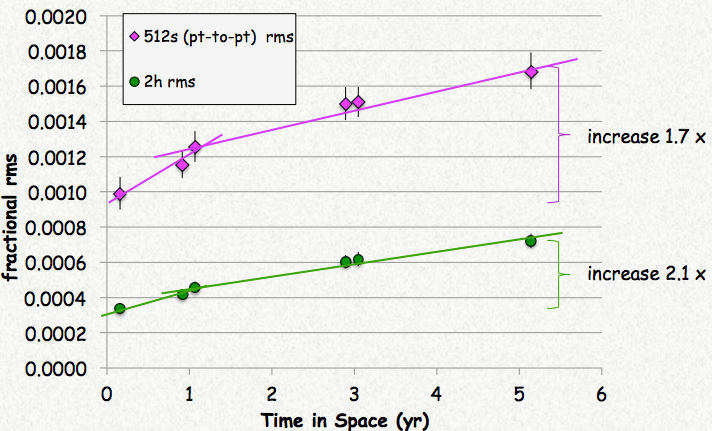}{\small{Noise evolution of the CCD against CoRoT time in space for a R=14 magnitude target. Point-to-point noise and the total noise on 2h timescales are shown in pink and green, respectively. The data points correspond, ordered in time, to the noise calculation of 56-days CoRoT data for IRa01, LRa01 (split in two sets of 56 days each), LRa03 (split in two sets of 56 days each) and LRa06. }}{}{0.9}

\figuremacroW{2.pdf}{\small{R magnitude against B-V magnitude for stars in IRa01, LRa01, LRa03 and LRa06}}{\small{The value of the magntitudes has been taken from Exodat. Blue dots show the stars classified as likely dwarfs, while orange dots are used for likely giants. As done in A09, a straight line running from B-V=1.3, 1.2, 1 and 1.2 at R=16 to B-V=0.7 at R=11 for IRa01, LRa01, LRa03 and LRa06 respectively, is used to separate dwarfs from giant stars. This figure is similar to Figure 4 of A09, except for the colours used.}}{1}

Each dot in Figures \ref{3.pdf} and \ref{4.pdf} represents the scatter measurement of a processed light curve over the run. Figure \ref{3.pdf} shows the point-to-point (512 sec) noise block together with that on long time scales (red) and the correlated noise over 2 hours (green). CoRoT's high-precision photometry is clearly demonstrated, with values for the scatter of about 10$^{-3}$ for point-to-point noise and down to 10$^{-4}$ for the correlated noise, for R magnitudes up to $\sim$13. The thick solid line is a lower envelope for the point-to-point noise obtained as a result of a least-squares fit to the lower quintil (or 20th percentile), which were determined for each bin of 0.5 magnitudes of the light curves with least scatter. The individual fits are shown in the central column of Table \ref{fits}. As expected, it does follow the same trend as the theoretical source photon noise (slope of 0.2, taken from A09), which is shown as a dashed line, but increases from $\sim$1.3 to $\sim$2 times the photon noise for IRa01 (0.2 years) and LRa06 ($\sim$5 years), respectively. \\

For the white noise on 2h time scales we have simply scaled the lower envelope of the point-to-point noise down by a factor of $\sqrt{14}$, shown as a dotted line. The correlated noise on 2h time scales (green dots), determined from solving Eq (3.1), increases with time on different runs; it is a factor of 0.8 times the 2h white uncorrelated noise for IRa01 and becomes significantly larger when reaching LRa06, of about 1.8 times the white uncorrelated noise. This is, red noise increases much more than white noise during the flight phase.\\

\begin{table}[h!]
\centering
\begin{tabular}{|c|c|c|} 
\hline
\textbf{Run} & \textbf{log(scatter)}& \textbf{log(scatter$\rm_{2h}$)} \\ 
\hline 

\textbf{IRa01} & (0.1892 $\pm$ 0.0015)R - 5.686$\pm$0.02 &(0.2063 $\pm$ 0.0014)R - 6.374$\pm$0.014 \\
\textbf{LRa01}& (0.2067 $\pm$ 0.0010)R - 5.821$\pm$0.014 & (0.2170 $\pm$ 0.0018)R - 6.377$\pm$0.014 \\
\textbf{LRa03}& (0.1975 $\pm$ 0.0009)R - 5.589$\pm$0.012 & (0.2078 $\pm$ 0.0010)R - 6.099$\pm$0.013 \\
\textbf{LRa06}& (0.1984 $\pm$ 0.0009)R - 5.606$\pm$0.013 & (0.2233 $\pm$ 0.0009)R - 6.275$\pm$0.012 \\
\hline
\end{tabular}
\caption[table]{\textbf{Least-squares fits to the lower quintil} - for the point-to-point scatter (thick solid line in Figure 3.3) and on 2h time scales (thick solid line in Figure 3.4)}
\label{fits} 
\end{table}

Stars are identified as dwarfs or giants following precepts from A09 (Figure \ref{2.pdf}) and the evolution of the total 2h scatter for these two groups in the four studied runs is shown in Figure \ref{4.pdf}. The white 2h noise is shown again as a dotted line for comparison.  A least-squares fit to the lower quintil of the total 2h noise of likely dwarf stars is performed (solid line in Figure \ref{4.pdf}). 

In order to represent the noise evolution of CoRoT, we decided to perform this same analysis, but in this case for an equal-sampled set of data. In this way, we take the data from the last 56 days of each run except for LRa01 and LRa03, from which we extract two datasets of 56 days starting from the end of the run, as these are of a longer duration. Therefore, we end up having 6 sets of data (one from IRa01, two from LRa01, two from LRa03 and one from LRa06) from which we compute the noise. Again, we fit the point-to-point and total 2h noise of dwarf stars, getting a very similar results to those obtained in Table \ref{fits}, reason why we do not show them here for clarity purposes. Instead, the result of these fits for R=14 magnitudes against CoRoT time in space is shown in Figure \ref{ageing}. Two striking conclusions can be extracted from the plot: first,  the point-to-point noise increases 1.7 times during the $\sim$5 years of observations, while the total noise on 2h timescales becomes 2.1 times higher than at the beginning of the mission, corresponding to a 3 times increase in correlated, assuming $\sigma\rm_{uncorrelated}$ and $\sigma\rm_{correlated}$ to be equal in the first run IRa01. On the other hand, two timescales of ageing seem to occur. During the first year there is a rapid ageing, or $burn-in$ phase, with a $\sim$30$\%$ noise increase. After that, during the following four years the noise still gets higher, but in a smaller rate of  $\sim$10--20$\%$ per year.

This ageing is likely due to an increase in the number of hot pixels detected in the CCDs. If a hot pixel is enclosed within the photometric aperture mask, an effect similar as the one shown in Figure \ref{1.pdf} (bottom) will be obtained, depending on the energy of the incoming proton. Thermal fluctuations will enhance the dark current level of the light curve, while its fluctuations with time will vary in different ways as the silicon lattice rearranges itself in stable configurations. These bright pixels can remain for years, which for later runs makes it more likely to find a hot pixel among the pixels used for the photometry of a given star.\\

The pre-launch specification of CoRoT's photometric performance was established at 7 $\times$ 10$^{-4}$ for a V=15.5 magnitude star for one hour of integration in order to be able to detect transit amplitudes for telluric planets \cite{bernardi01}. As only white noise was assumed, we can translate this threshold to a scatter of $\frac{7 \times 10^{-4}}{\sqrt{2}}$ = 5.5 $\times$ 10$^{-4}$ in 2h for V=15.5. This magnitude corresponds to R=15 for a CoRoT dwarf target (A09). The scatter obtained from Table \ref{fits} for R=15 targets on 2h time scales is 5.2 $\times$ 10$^{-4}$ for IRa01(95$\%$ of the specified noise), 7.55 $\times$ 10$^{-4}$ for LRa01, 1.05 $\times$ 10$^{-3}$ for LRa03 and 1.17 $\times$ 10$^{-3}$ for LRa06. This clearly confirms the degradation of the photometric performance already pointed out by A09 for the first 14 months of data.\\

The 2.1 times increase in the 2h scatter has an impact on the potential detectability of planetary transits. The stellar flux decrease, or the depth of a transit, is given by:
\begin{equation}
\frac{\Delta F}{F} \propto \left( \frac{R_{p}}{R_{s}} \right)^{2}
\end{equation}

where $R_{p}$ is the radius of the planet and $R_{s}$ the radius of its host star. As a very rough estimation, if we wish to have a signal to noise ratio of (let's say) 3, only exoplanets with a stellar flux decrease of at least 3 $\times$ scatter$_{2h}$ would be detected. This is translated to detectable planets of radii 3.9\% and 5.9\% of that of its host star for IRa01 and LRa06, respectively. Assuming a host star like our Sun, CoRoT would potentially detect planets with $\sim$3.9$\times$R$\rm_{Earth}$ for IRa01 and $\sim$5.9$\times$R$\rm_{Earth}$ for LRa06. For a fainter star with half the radius of the Sun, planetary radii go down to 2$\times$R$_{Earth}$ for IRa01 and 3$\times$R$_{Earth}$ for LRa06. This increase in noise on transi time scales is certainly also a relevant factor for the lower detection yields in the later runs (Deleuil et al., in prep).

\section{Zero point and Instrumental Throughput}

The difference between the photons collected by CoRoT and the digital counts that we finally detect depends on the transmission of the telescope, the quantum efficiency of the detector and the gain value when converting the analogue signal of the electrons to counts. This total throughput influences the photometric zero point present in the transformation equation between the CoRoT system and the Johnson -Kron-Cousins standard magnitudes (Leach 1987) (as used by Landolt 1992):

\begin{equation}
R_{s} = Z_{p} + C(V_{s}-R_{s}) + W_{CoRoT}
\end{equation}
where the subindex \textit{s} indicates magnitude in the standard system, W$\rm_{CoRoT}$ is the magnitude in white CoRoT \textit{colour} calculated as - 2.5log(counts), Z$_{p}$ is the photometric zero point and C is a colour-correction constant. As CoRoT is a space-based instrument, there is no need for an atmospheric extinction coefficient. For the counts, the light curves' median value was taken after all the corrections were performed. \\

The zero point is the magnitude at which a count of 1 is received. A bigger zero point value means a higher sensitivity, this is, more electrons are produced as output from a given source. Figure \ref{zpp} shows the zero point values (given as R - White$_{CoRoT}$) for all the light curves in the four studied runs, assuming no colour correction is made (e.g., we used C=0 since no consistent V$_{S}$ - R$_{S}$ colour dependency could be detected). The median zero point is Z$\rm_{p}$ = 26.723 $\pm$ 0.003, 26.695 $\pm$ 0.002,  26.700 $\pm$ 0.004 and 26.664 $\pm$ 0.003 magnitudes for IRa01, LRa01, LRa03 and LRa06, respectively (see Figure \ref{zpp}). As it remains within 0.05 magnitudes ($\sim$5$\%$ in flux units), we can state that there is not a significant change in the CoRoT's transmission, quantum efficiency and gain. 

\figuremacroW{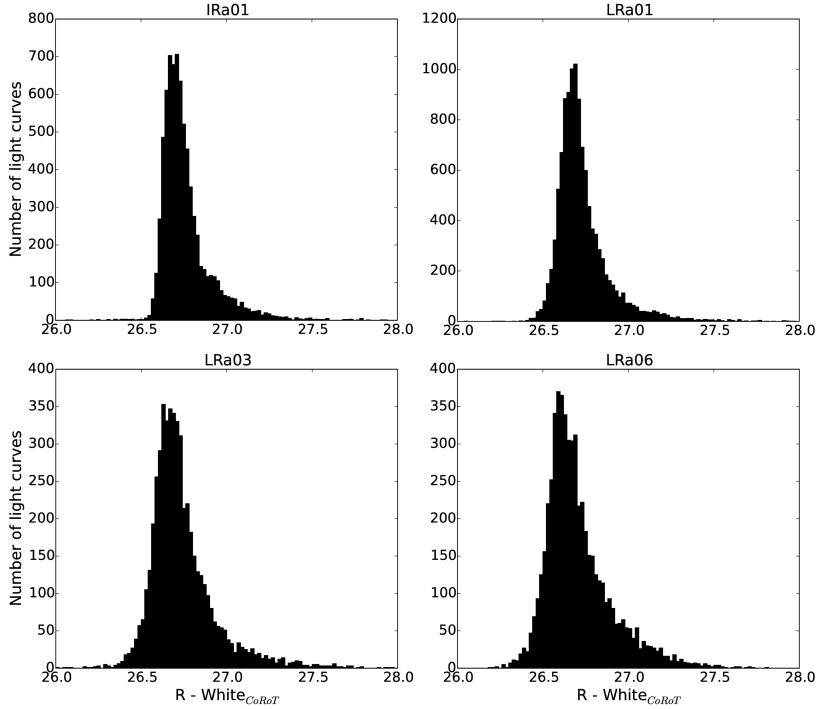}{\footnotesize{Number of light curves versus the zero point value (given as R - White$\rm_{CoRoT}$) for IRa01, LRa01, LRa03 and LRa06}}{\footnotesize{R magnitudes are taken from the Exodat catalogue, while  White$\rm_{CoRoT}$ = -2.5log(counts)}}{0.8}


\chapter{CONCLUSION}


\ifpdf
    \graphicspath{{4/figures/PNG/}{4/figures/PDF/}{4/figures/}}
\else
    \graphicspath{{4/figures/EPS/}{4/figures/}}
\fi


We have confirmed the photometric degradation of CoRoT CCDs during the whole flight phase (2007-2012), which was pointed out by A09 for the first 14 months of data. Our study of the noise level of CoRoT light curves is based on runs IRa01 (2007) and LRa01 (2008), also studied by A09, but it has been extended to LRa03 (2010) and LRa06 (2012). We find a receding photometric performance from the pre-launch specification on transit time scales (2h). These specifications were only fulfilled in the first run analyzed (IRa01), for which the 2h noise for an R=15.5 mag target amounts to 95$\%$ of the specified noise. Our results match those obtained by A09 for IRa01 and LRa01, while the degradation continues for LRa01 and LRa06. The total noise on 2h time scales increases by a factor 2.1 between the first and the last run analysed; that is, over a span of about 5 years. This corresponds to an increase of $\sim$15$\%$ per year. However, we find the strongest degradation to be produced in the first year ($\sim$30$\%$ noise increase). Also on 2h time scales, the correlated noise becomes more important than white uncorrelated noise for the two last studied runs, with the uncorrelated noise (point-to-point noise on 8 min timescale) increasing only by a factor of 1.7 across the 5 year span.\\

The point-to-point noise, although low, is about three times bigger than the source photon noise. As expected, this noise increases with decreasing stellar brightness in the same manner as the photon noise for faint stars. We find here a difference with the results reported by A09, who found a steeper slope for the point-to-point noise versus magnitude of 0.25, whereas we find a slope close to 0.20, which is the one expected for noise dominated by Poisson statistics. This noise on short timescales (512 sec / 8 min) increases slightly less.\\

The increases in noise with age seems to affect all or nearly all CoRoT apertures, with at least $\sim$5$\%$ of the pixels having got degraded during the mission lifetime. The data taken during the Southern Atlantic Anomaly (SAA) crossings is strongly degraded, so we consider it the main source of radiation damage from hot pixel events, dominanting the ageing effect on CoRoT's performance with time. SAA will also affect CHEOPS, also on a polar orbit at slightly less altitude. Finally, we find a nearly constant zeropoint that excludes a relevant decrease in the instrument efficiency, as was observed on WFC3/HST after 3 years of operation \cite{baggett01}. \\




\begin{acknowledgements}      

I deeply thank my supervisor, Dr. Hans J. Deeg, for guiding and helping me from the early stages to the final writing of the project. His advices and corrections have been of vital importance for the completion of this MSc thesis.

\end{acknowledgements}




\begin{scriptsize} 

\bibliographystyle{plainnat} 
\renewcommand{\bibname}{References}  
\bibliography{references} 

\end{scriptsize}








\end{document}